\documentclass[preprints,article,accept,moreauthors,pdftex]{Definitions/mdpi} 

\usepackage{bm}
\usepackage{bbm}
\usepackage{listings}
\usepackage{subfig}
\usepackage{amsmath}
\usepackage{amssymb}
\usepackage{amsthm}
\usepackage{booktabs}
\usepackage{hyperref}
\usepackage{fourier}
\usepackage[T1]{fontenc}
\usepackage[]{avant}
\usepackage{longtable}
\usepackage[ruled, lined, linesnumbered, commentsnumbered, longend]{algorithm2e}

\hypersetup{colorlinks,citecolor=blue}

\firstpage{1} 
\pubvolume{xx}
\issuenum{1}
\articlenumber{1}
\pubyear{2023}
\copyrightyear{2023}
\history{}

\newcommand*\diff{\mathop{}\!\kern0pt\mathrm{d}}

\newcommand{\sinc}{\mathrm{sinc}}

\Title{Notes on the SWIFT method based on Shannon Wavelets for Option Pricing - Revisited}
\Author{Fabien Le Floc'h}

\AuthorNames{Fabien Le Floc'h}

\address{fabien@2ipi.com}
\abstract{This note revisits the SWIFT method based on Shannon wavelets to price European options under models with a known characteristic function in 2023. In particular, it discusses some possible improvements and exposes some concrete drawbacks of the method.}

\begin{document}
	
	\section{Introduction}
	\citet{ortiz2016highly} describe a new method, called SWIFT, based on Shannon wavelets, to price European options  for models with a known characteristic function. We will apply the technique mostly to the Heston stochastic volatility model \citep{heston1993closed}. \citet{lefloch2020notes} explored alternative implementations of the SWIFT method, to conclude that the use of fast Fourier transform (FFT) was key, and suggested that the use of the trapezoidal rule to compute the density coefficients was more accurate than the use of an expansion of the Vieta formula as in \citep{ortiz2016highly}.
	
	We revisit here which quadrature is more appropriate for the computation of the density coefficients and of the payoff coefficients. We also pay more attention to the choice of SWIFT parameters, as the various publications on SWIFT propose slight variations. Finally we show through concrete examples that the method suffers from many of the same drawbacks as the COS method.
	
	\section{Quick summary of the SWIFT method}
	With the SWIFT method, the price at time $t$ of a Vanilla Put option of maturity $T$ and log-moneyness $x=\ln\frac{F}{K}$, with $K$ the strike and forward $F$ reads\footnote{We adopted the simplification of \citet{maree2017pricing} and \citet{lefloch2020notes} where the sum is symmetric over $2\kappa$ elements, which is optimal from a FFT perspective. It also makes sense from an implementation perspective. A generalization to a non-centered interval is trivial.}
	\begin{equation}
		v(x,t) = B(t,T) \sum_{k=1-\kappa}^{\kappa} c_{m,k} V_{m,k}\,,
	\end{equation}
	where 
	\begin{align}
		c_{m,k} = \left\langle f | \phi_{m,k} \right\rangle = \int_{\mathbb{R}} f(x) \phi_{m,k}(x) \diff x &\,,\quad		V_{m,k} = \int_{I_m} v(y,T)\phi_{m,k}(y) \diff y\,,\label{eqn:vmk}\\
		\phi_{m,k}(x) = 2^{\frac{m}{2}}\phi\left(2^m x - k\right) &\,, \quad 		\phi(x) = \sinc(x) =  \frac{\sin \pi x}{\pi x}\,,
	\end{align}
	and $(m, \kappa) \in \mathbb{N}\times \mathbb{N}^{\star} $, suitably chosen, $f$ the probability density function and $v(y,T)$ is the payoff at maturity with $y=\ln\frac{F(T,T)}{K}$, that is  $v(y,T)=K|1-e^y|^+$ for a vanilla Put option.

	For the models we are interested in, $f$ is not known directly but its Fourier transform is, we have $\hat{f}(\omega,x) = \mathbb{E}\left[e^{-i\omega X_T}\right]=\psi(-\omega,x)$ where $\psi$ is the characteristic function. For L\'evy models, it may be further decomposed into$ \hat{f}(\omega,x) =\hat{f}(\omega,0) e^{-i\omega x}$. We may then use simply the notation $\hat{f}(\omega) := \hat{f}(\omega,0)$.
	
	In \citep{ortiz2016highly}, the coefficients $c_{m,k}$ and $V_{m,k}$ are computed using an approximation based on Vieta formula for the cardinal sinus:
	\begin{equation}
	\sinc(t) \approx \frac{1}{J} \sum_{j=1}^J  \cos(\omega_j t)\,,
\end{equation}
with $\omega_j = \frac{\pi}{J}\left(j-\frac{1}{2}\right)$. 
The coefficients $c_{m,k}$ and $V_{m,k}$ read
\begin{align}
	c_{m,k} &= \frac{2^{\frac{m}{2}}}{J} \sum_{j=1}^J \Re\left\{ \hat{f}(\omega_j 2^m, x) e^{i \omega_j k}\right\}\,,\\
	V_{m,k} &= K \frac{2^{\frac{m}{2}}}{J} \sum_{j=1}^J \Re\left\{ e^{-i \omega_j k} \int_{0}^c (1-e^y) e^{i \omega_j 2^m y} \diff y\right\}\,,
\end{align}
where $c = \frac{\kappa}{2^m}$.
	
	\citet{lefloch2020notes} proposes another formula, centered on the forward, such that the characteristic function is only evaluated in the center ($x=0$):
		\begin{align}
	V_{m,k} &= 	\frac{F 2^{\frac{m}{2}}}{J} \sum_{j=1}^{J} \int_{-c}^{-x} \left(e^{-x} - e^y\right) \cos \left(\omega_j \left(2^m y -k\right) \right) \diff y\nonumber \\
	&-\frac{\pi F 2^{\frac{m}{2}}  }{24 J^2} \int_{-c}^{-x} \left(2^m y -k\right)\left(e^{-x} - e^y\right) \sin \left(\pi \left(2^m y -k\right) \right) \diff y\,.\label{eqn:vmk_maclaurin_second}
\end{align}
The $V_{m,k}$ become dependent on $x$ while $c_{m,k}$ are the same as in the standard SWIFT method but always evaluated at $x=0$.
	It relies on the second Euler-Maclaurin formula (including the correction term with the first derivative) for increased accuracy.
	The second Euler-Maclaurin formula for $\sinc$ reads \citep{belov2014coefficients}
	\begin{equation}
		\frac{\sin(\pi x)}{\pi x} \approx \frac{1}{J} \sum_{j=1}^{J} \cos \left(\omega_j x \right) - \frac{\pi x}{24 J^2} \sin (\pi x)\,.
	\end{equation}
	It also is a more direct way to establish the accuracy of the approximation.  The identity
	\begin{equation}\sinc(x) = \frac{1}{\pi}\int _0^\pi \cos(tx) \diff t \label{eqn:sinc_integral_repr} \end{equation}
	may also be discretized with various quadrature rules, which opens up to yet other kind of approximations. The mid-point rule applied to the above integral is equivalent to the  approximation based on the Vieta expansion \citep{maree2017pricing}.

	\section{Choice of SWIFT parameters}
	\citet{maree2017pricing} describe a precise procedure to select firstly the scale of the wavelet $m$ from the characteristic function at the end-points:
	\begin{equation}
		\epsilon_m = \frac{(2^m \pi)^{1-\nu}}{2\pi\nu T}\left(|\hat{f}(-2^m\pi,x)| +|\hat{f}(2^m\pi,x)| \right) \label{eqn:epsilon_m_maree}
	\end{equation}
	where $\nu$ is such that $\exists (d,C) \in {\mathbb{R}^{+\star}}^2, |\hat{f}(\omega, x)| \leq C e^{-d \tau |\omega|^\nu}$. In particular for the Heston model, we have $\nu = 1$. The algorithm to find $m$ is to increment $m$ by one until the threshold $\epsilon_m$ is reached.

	Starting from $c=|c_1| + L\sqrt{|c_2|+\sqrt{|c_4|}}$, the  interval boundary is set to $\kappa=\lceil 2^m c \rceil$ and increased such that  the total density sums to one up to $\epsilon_f$ using the following equation
	\begin{equation}
		\epsilon_f = \left| 1 - 2^{-\frac{m}{2}} \sum_{k=1-\kappa}^{\kappa} c_{m,k}\right|\,. \label{eqn:swift_density_eps}
	\end{equation}
	Previous coefficients may\footnote{We will see that it is  true if $J$ does not change or if the trapezoidal rule is used.} be reused in this latter iterative method. Then $2^{\log_2(J)} = \lceil \pi \kappa \rceil$ or equivalently $\log_2(J)= \lceil \log_2(\pi \kappa) \rceil$. 
	
	\citet{romo2021swift} describe a slightly different procedure and updates $m$ according to the total density criterion (Equation \ref{eqn:swift_density_eps}) . $\kappa, J$ are set the same way as above. This idea does always not work as the error does not decrease up to the tolerance if the estimate of $c$ is not good enough. For example let us consider the Heston parameters $\kappa=4, \theta= 0.25, \sigma= 1, \rho=-0.5, v_0= 0.01, T=0.01$. Starting with a smallish $L=4$ and $m=8$ leads to the iteration of Table \ref{tbl:iteration_romo}. The estimate of $m$ from Equation \ref{eqn:epsilon_m_maree} is 9 with a tolerance of $10^{-8}$.
	\begin{table}[h]
		\caption{Estimate $\epsilon_f$ in the iterative procedure of \citet{romo2021swift} (left) and \citet{leitao2018swift} (right) starting with a too small $L$. \label{tbl:iteration_romo}}\centering{
		\begin{tabular}{lllr}\toprule
			$m$ & $\kappa$ & $\log_2 J$ & $\epsilon_f$\\\midrule
			8 & 18 &6 & 7.130920268738627e-5\\
	9 & 35 & 7& 6.961299004415444e-5\\
	10 & 69 & 8 &6.918898569008292e-5\\
	11 & 138 &9 &6.379632495345788e-5\\
	\multicolumn{3}{c}{...}\\
	21 & 141197 & 19 & 5.941240610984888e-5\\
	22 & 282394 & 20 & 5.941009234999850e-5\\\bottomrule
			\end{tabular}
			\begin{tabular}{lllr}\toprule
			$m$ & $\kappa$ & $\log_2 J$ & $\epsilon_f$\\\midrule
			8 & 18 &6 & 7.130920268738627e-5\\
			8 & 21 & 7& 1.0255976256812183e-5\\
			8 & 25 & 7 &7.089187974429478e-7\\
			8 & 35 &7 &8.968050746460676e-10\\
			\multicolumn{3}{c}{...}\\
		8 & 49921 & 18 &3.3306690738754696e-16\\
		8 & 59366 & 18 & 0.0	\\\bottomrule
			\end{tabular}}
	\end{table}
	In particular, the estimate does not decrease below $5.9\cdot10^{-5}$ even for very large $m$ or $J$ and thus crucially depends on a good estimate of $L$.
	
	\citet{leitao2018swift} offer another variation. The tolerance criteria for $m$ is independent on the time to maturity
		\begin{equation}
			\epsilon_m = \frac{1}{2\pi}\left(|\hat{f}(-2^m\pi,x)| +|\hat{f}(2^m\pi,x)| \right)
		\end{equation} 
	 and $c$ itself is updated according to the total density criteria (Equation \ref{eqn:swift_density_eps}) while  $\kappa$ and $J$ are recomputed from the new $c$ in the iterative procedure. This is more in line with \citep{ortiz2016highly}, where $c=\frac{\kappa-1}{2^m}$ is used in the calculation of the put payoff coefficient. The estimate of $m$ is still somewhat critical to be able to reach the accuracy set for $\epsilon_f$: if $\epsilon_m$ is set to be too large, $m$ may be too small to reach a the accuracy set for $\epsilon_f$.  This may explain why \citet{romo2021swift} proposed to circumvent $\epsilon_m$ altogether (at the cost of a strong sensitivity on the choice of $L$ as explained above). In practice, we found the estimate of $m$ to be too conservative when using similar $\epsilon_m$ and $\epsilon_f$ thresholds, so the procedure often works, but may lead to a 2x or 4x increase of the number of points used.
	 
	The range $[-c, c]$ is adjusted using the minimum and maximum quoted log-moneyness to $[x_{\min}-c,x_{\max}+c]$. In this note, we don't apply this adjustment but use the payoff formula centered on the forward of \citet{lefloch2020notes}. The interval adjustment along with the strike based formula would not change any of the observations made in this note.
	
	The choice $\log_2(J) = \lceil \log_2(\pi \kappa) \rceil$ is motivated by the range allowed in the series which constitutes the  upper bound of the error in the sinc function approximation in  \citep{ortiz2016highly}. In reality, the more direct upper bound  from the Euler-Maclaurin expansion (Equation \ref{eqn:vmk_maclaurin_second}) is much stricter and the choice $\log_2(J) = \lceil \log_2( \kappa) \rceil$ is enough for the FFT algorithm on $N=2J$ to generate the indices in the interval $\{-\kappa+1, ..., \kappa \}$.
	\begin{proof}
		According to \citet{belov2014coefficients}, the second Euler-Maclaurin expansion of sinc reads
		\begin{align*}		\sinc(x) = \frac{1}{J}\sum_{j=1}^J \cos (\omega_j x) +  \frac{1}{\pi}\sum_{k=1}^\infty (-1)^{k+1} b_k  \left(\frac{\pi}{J}\right)^{2k} x^{2k-1} (-1)^k \sin(\pi x)\,,
		\end{align*}		
	with $b_k > 0$ such that $b_k <  0.08 b_{k-1} $ 	and $b_1=\frac{1}{24}$.
	We thus have
	\begin{align*}
		\left|\sinc(x) - \frac{1}{J}\sum_{j=1}^J \cos (\omega_j x) \right| \leq \frac{b_1}{J}\sum_{k=1}^\infty 0.08^{k-1}  \left(\frac{\pi x}{J}\right)^{2k-1}\,,
	\end{align*}
	or equivalently
	\begin{align}
		\left|\sinc(x) - \frac{1}{J}\sum_{j=1}^J \cos (\omega_j x) \right| \leq \frac{b_1 \pi x}{J^2}\sum_{k=1}^\infty   \left( \left(\frac{\pi x}{J}\right)^2 0.08\right)^{k-1}\,. \label{eqn:sinc_upperbound}
	\end{align}
	The series on the right hand side converges for $\left|\left(\frac{\pi x}{J}\right)^2 0.08\right| < 1$, which implies $J > \pi \sqrt{0.08}|x|$. Furthermore $\pi \sqrt{0.08} \leq 0.89$ and thus $J = \lceil\ |x|\rceil$ is sufficient. Furthermore, the error is then bounded by $\frac{b_1 \pi x}{J^2} \frac{1}{1-\frac{0.08\pi^2   x^2}{J^2}}$.
	
	 For the trapezoidal rule (first Euler-Maclaurin formula), the error bound would be very similar, but replacing $b_1$ by $a_1 = 2b_1$.
	\end{proof} It may be however worth using $\lceil\log_2(\kappa)\rceil +1$ to guarantee a reasonably good accuracy up to the index $k=\kappa$.
	
	We will use the procedure of \citet{leitao2018swift} in our numerical examples, because it is much less sensitive to the choice of truncation interval defined by $L, c$, a clear improvement\footnote{the procedure of \citet{romo2021swift} with a large $L$ may be a  competitive choice.} over the COS method of \citet{fang2008novel}, unless stated otherwise.
	
	\section{Vieta, Mid-points or Trapezoidal?}
	\subsection{For the payoff}
	Truncated to the interval $[-c,c]$ the payoff coefficients read
	\begin{equation}
		 V_{m,k} =  2^{\frac{m}{2}} K e^x \int_{-c}^{-x} \left(e^{-x} - e^y\right) \frac{\sin \left(\pi \left(2^m y -k\right)\right)}{\pi \left(2^m y -k\right)}\diff y\,. \label{eqn:swift_payoff_new}
	\end{equation}
	The integral may be approximated directly by some quadrature rule, such as the mid-point rule, the trapezoidal rule, or higher order quadrature rules such as Simpson or Boole. 	
	Applied directly to Equation \ref{eqn:swift_payoff_new}, the mid-point quadrature  is not the same as the approximation based on the Vieta expansion, since the mid-points are applied to the integrand, and thus the sinc function instead of the cosine function.
	
	Alternatively, we may integrate exactly an approximation of the sinc function. Equation \ref{eqn:vmk_maclaurin_second} uses the second Euler-Maclaurin expansion with first order derivatives.  Excluding the second-term, we have the mid-point rule applied to the sinc function integral representation (Equation \ref{eqn:sinc_integral_repr}). In similar fashion we may apply the first Euler-Maclaurin expansion to obtain
		\begin{align}
	V_{m,k} &= 	\frac{2^{\frac{m}{2}}F }{J}  \int_{-c}^{-x} \left(e^{-x} - e^y\right) \left[\frac{1}{2} +\frac{1}{2}\cos \left(\pi \left(2^m y -k\right)\right) + \sum_{j=1}^{J-1}  \cos \left(\frac{\pi j}{J} \left(2^m y -k\right) \right)\right] \diff y\nonumber \\
	&+\frac{\pi 2^{\frac{m}{2}}  F}{12 J^2} \int_{-c}^{-x} \left(2^m y -k\right)\left(e^{-x} - e^y\right) \sin \left(\pi \left(2^m y -k\right) \right) \diff y\,.\label{eqn:vmk_maclaurin_first}
\end{align}
	Excluding the second term, it corresponds to the trapezoidal rule applied to the sinc function integral representation. We may pursue the idea with the Simpson rule. The rule involving $2J+1$ terms corresponds to $2/3$ of the mid-point rule plus $1/3$ of the trapezoidal rule, and the first derivative Euler-Maclaurin correction is just zero.
	
	Equation \ref{eqn:vmk_maclaurin_first} may be computed via the FFT as we can rewrite it in the following form:
	\begin{align}
		V_{m,k} &= 	\frac{2^{\frac{m}{2}}F }{J} \Re\left\{ \sum_{j=0}^{J}  w_j    e^{-i\pi\frac{kj}{J}} \left[ g_j(-x)-g_j(-c) \right] \right\}\nonumber\\
		&+\frac{(-1)^{k}\pi 2^{\frac{m}{2}}  F}{12 J^2} \Im\left\{  -k  \left[ g_J(-x)-g_J(-c)\right] + 2^m \left[ -x g_J(-x) + c g_J(-c)\right] - 2^m\left[  -\frac{e^{-x+i p_J z}}{ p_J^2} - \frac{e^{z+i p_J z}}{(ip_J+1)^2} \right]_{z=-c}^{z=-x}   \right\}
	\end{align}
	with $p_j = \frac{2^m \pi j}{J}$ and $g_j(z) = \frac{e^{-x+i p_j z}}{i p_j} - \frac{e^{z+i p_j z}}{i p_j+1}$ for $j \geq 1$ and $g_0(z)= z-e^z$, $w_j=1$ for $j=1,...,J-1$ and $w_0=w_J=\frac{1}{2}$. 
	
\begin{figure}[h]
		\centering{
			\subfloat[Value]{
			\includegraphics[width=0.48\textwidth]{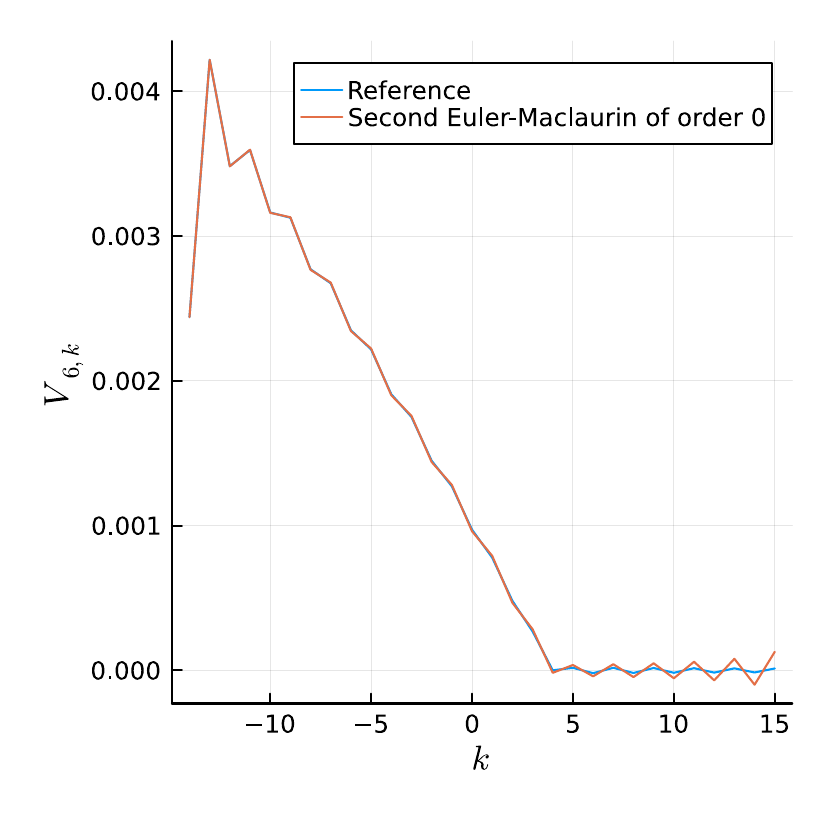}}
			\subfloat[Absolute error]{
			\includegraphics[width=0.48\textwidth]{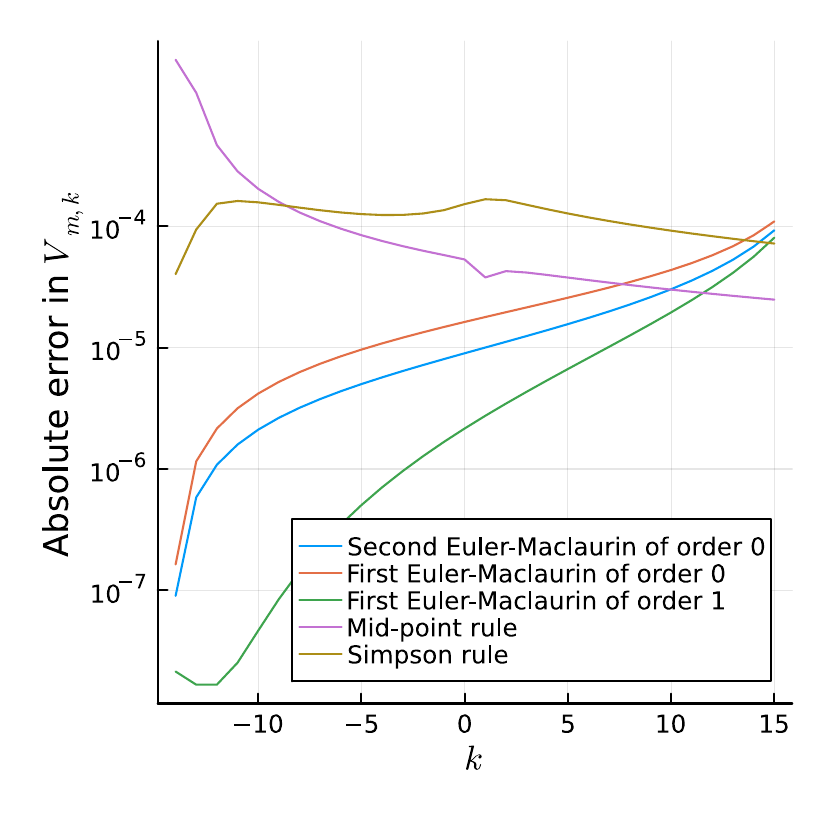}}
		}
			\caption{\label{fig:v6k_secondem0_set2}Value and error in the coefficients $V_{6,k}$ for the option of strike $K=1.064$, using $m=6, L=8, J=2^4$.}
	\end{figure}

	We apply the variety of approximations to the Set 2 of \citet{lefloch2020notes}. In Figure \ref{fig:v6k_secondem0_set2}, we plot the $V_{m,k}$ and the absolute  error in $V_{m,k}$ for the out-of-the-money put option with strike $K=1.064$.  With $\epsilon_m = 10^{-8}$,  the procedure of \citet{leitao2018swift} leads to $m=7$, and with $\epsilon_m=10^{-4}$, we obtain $m=6$. We use the latter for a better comparison with the results in \citep{lefloch2020notes}. We then use $\epsilon_f = 10^{-4}$ which leads to $J=2^4$. For the calculation of the density coefficients $c_{m,k}$ we use a different, much larger $J$, in such that the error in their values is negligible for our purpose. This is a relatively small number of points, and we may expect to see the impact of the various approximations more clearly.
	
	We observe that the quadrature rules based the discretization of Equation \ref{eqn:swift_payoff_new} are much less accurate than the Euler-Maclaurin expansions of the sinc function, except close to the boundary $k=\kappa$. The Simpson quadrature does not improve over the mid-point rule, likely because the integrand is strongly oscillating. We also notice that the first derivative correction of the Euler-Maclaurin expansion improves the accuracy of the $V_{m,k}$ significantly.
	
\begin{figure}[h]
		\centering{
			\subfloat[Density coefficients $c_{6,k}$.]{
				\includegraphics[width=0.48\textwidth]{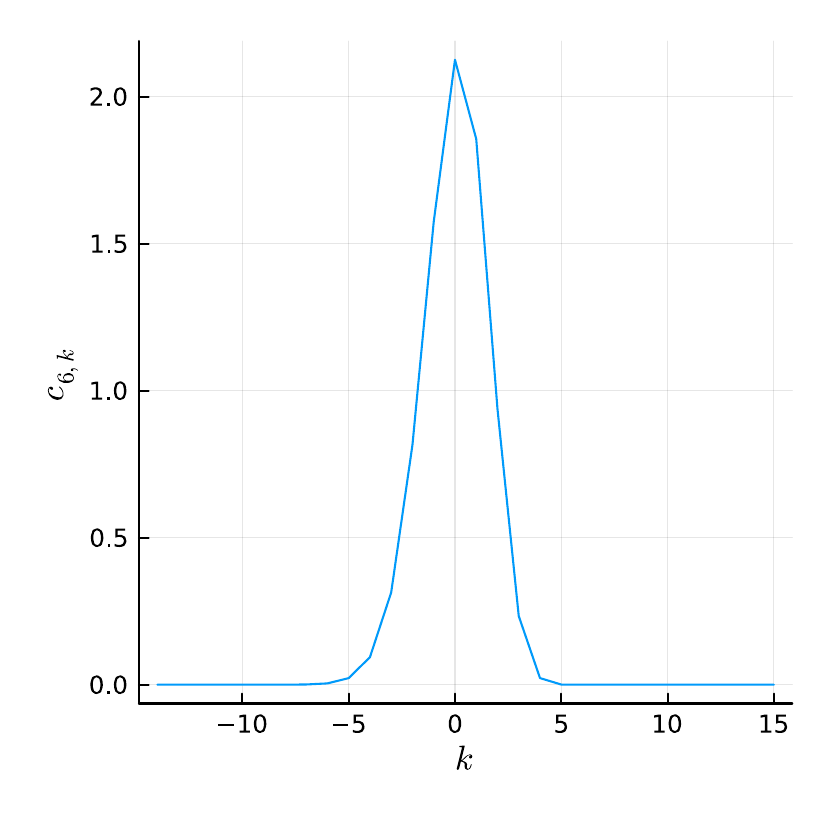}}
			\subfloat[Error in price]{
				\includegraphics[width=0.48\textwidth]{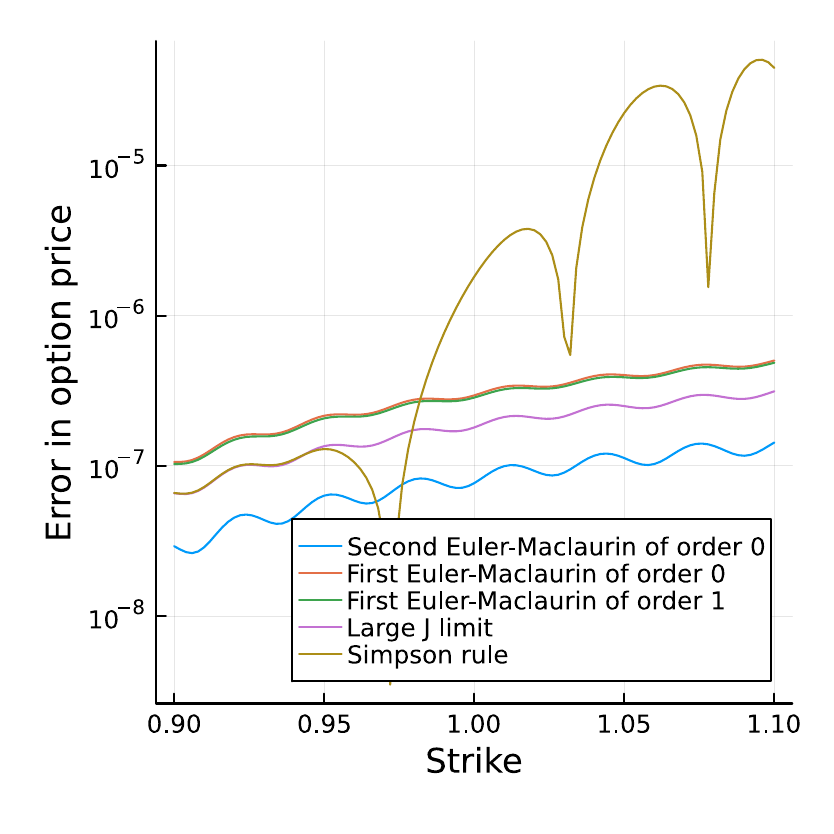}}
		}
		\caption{\label{fig:price_error_set2}Density coefficients and in vanilla option prices for a range of strikes, using $m=6, L=8, J=2^4$.}
	\end{figure}

	
	In terms of option prices, the quadrature rules lead to a much higher error than the Euler-Maclaurin expansions (Figure \ref{fig:price_error_set2}). There is however very little difference between the various Euler-Maclaurin expansions. In particular, the first derivative correction does not significantly improve the accuracy in the option price.
	Only if we let $J$ vary as a function of $k$ (see Theorem 1 of \citet{ortiz2016highly}), in which case the FFT is not applicable, and we see a more significant improvement with the Euler-Maclaurin correction.  There is however no good reason to use a varying $J$ since it is slower (due to the lack of FFT) and less accurate (due to the use of a smaller number of points).
	We also note that a much smaller error is reached if we use $m=7$ instead of $m=6$ but the conclusions stay the same.
	
	In all the combinations of Heston parameters and SWIFT parameters sets, we tried, the accuracy of the Vieta based approximation is as good as the approximations with Euler-MacLaurin corrections: there is no practical gain in spite of the increased accuracy of the $V_{m,k}$.

	\subsection{For the density}
	The mid-point rule applied to Equation \ref{eqn:density_parseval} is equivalent to the use of the Vieta based expansion for sinc, as proved in \citep{lefloch2020notes}.	It was found in the latter note that the trapezoidal rule was more accurate than the mid-points. This is not correct for two reasons:
	\begin{itemize}
		\item the rule was applied without  the correct weight at the right end. In reality a weight of $1/2$ should be applied there because the trapezoidal rule is applied to the Parseval identity.
		\item it was more accurate for the specific choices of $m$ in the note, but not for others or on other Heston parameters.
	\end{itemize}
	
	\begin{table}[h]
		\caption{Error in the out-of-the-money option prices for different approximations of the coefficients. The strike ranges are $[0.25\cdot 10^6, 4.0 \cdot 10^6]$  and $[0.9\cdot 10^6,1.1\cdot 10^6]$ respectively for Set 1 and 2. SEM0 stands for the second Euler-Maclaurin formula, FEM0/FEM1 for the first Euler-Maclaurin without/with first derivative correction. The initialization procedure leads to $\kappa=1568$, $\kappa=6$ and $\kappa=22$ for the three cases considered.\label{tbl:set12}}\centering{
		\begin{tabular}{llrrrrrr}\toprule
			Density & Payoff & \multicolumn{2}{c}{Set 1 ($m=8,J=2^{11}$) }& \multicolumn{2}{c}{Set 2 ($m=4,J=2^{3}$) } & \multicolumn{2}{c}{Set 2 ($m=6,J=2^{5}$) }\\ \cmidrule(lr){3-4}\cmidrule(lr){5-6}\cmidrule(lr){7-8}
			& & RMSE & MAE & RMSE & MAE & RMSE & MAE \\
			  SEM0 & SEM0 & 0.892& 1.490 & 2039 & 3233& 0.207 & 0.290\\
			 SEM0 & FEM0 &  0.906 &1.514& \textbf{1261} & \textbf{2142} & 0.252 & 0.350\\
			 SEM0 & FEM1& 0.900 & 1.504 & 1758 & 2824 &0.245 & 0.341\\
			 FEM0 & SEM0 & \textbf{0.812}& \textbf{1.357} & 2210 & 3456& \textbf{0.183} & \textbf{0.258}\\
			 FEM0 & FEM0 & 0.822 &  1.374 &1445 & 2351& 0.220 & 0.308\\
			 FEM0 & FEM1 & 0.818 &1.367 & 1937 & 3062& 0.215 & 0.301\\
			 FEM1 & SEM0 & 0.858 & 1.434 & 2127 & 3349 &0.200 & 0.281\\
			 FEM1 & FEM0 &  0.870 & 1.455 &1354 & 2246 &0.242 & 0.337\\
			 FEM1 & FEM1 &  0.865 & 1.446 & 1849 & 2948 &0.236 &0.329 \\\bottomrule
		\end{tabular}}
	\end{table}
	
	In reality, as shown in Table \ref{tbl:set12} which reproduces the example of \citet{lefloch2020notes}, the accuracy of the two rules is essentially the same.
	
	\begin{figure}[h]
		\centering{
			\subfloat[Error in density coefficients $c_{6,k}$.]{
				\includegraphics[width=0.48\textwidth]{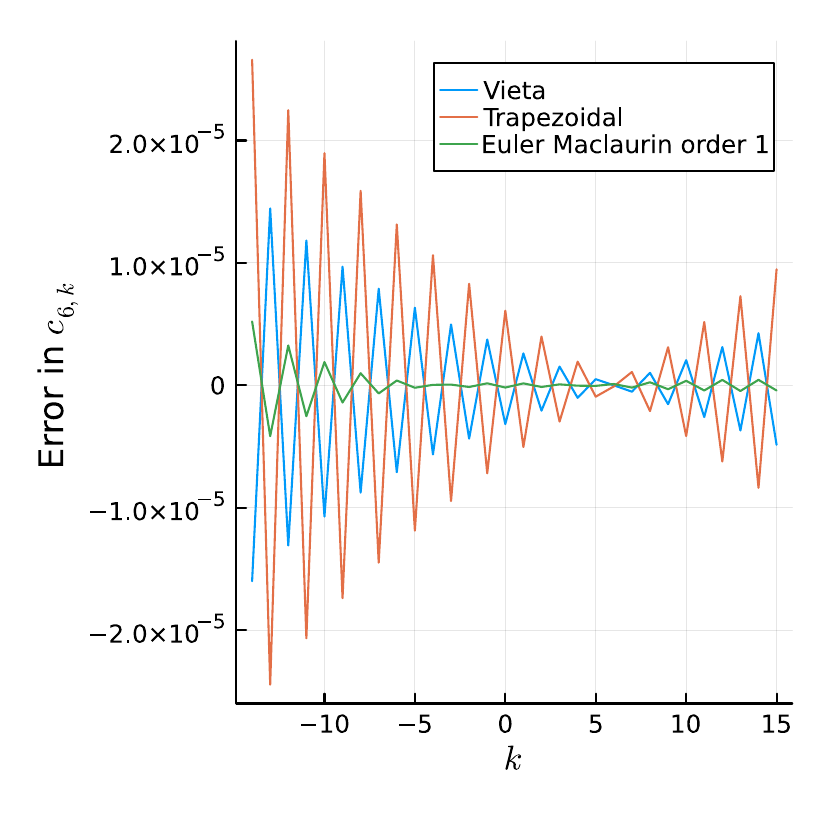}}
			\subfloat[Error in price]{
				\includegraphics[width=0.48\textwidth]{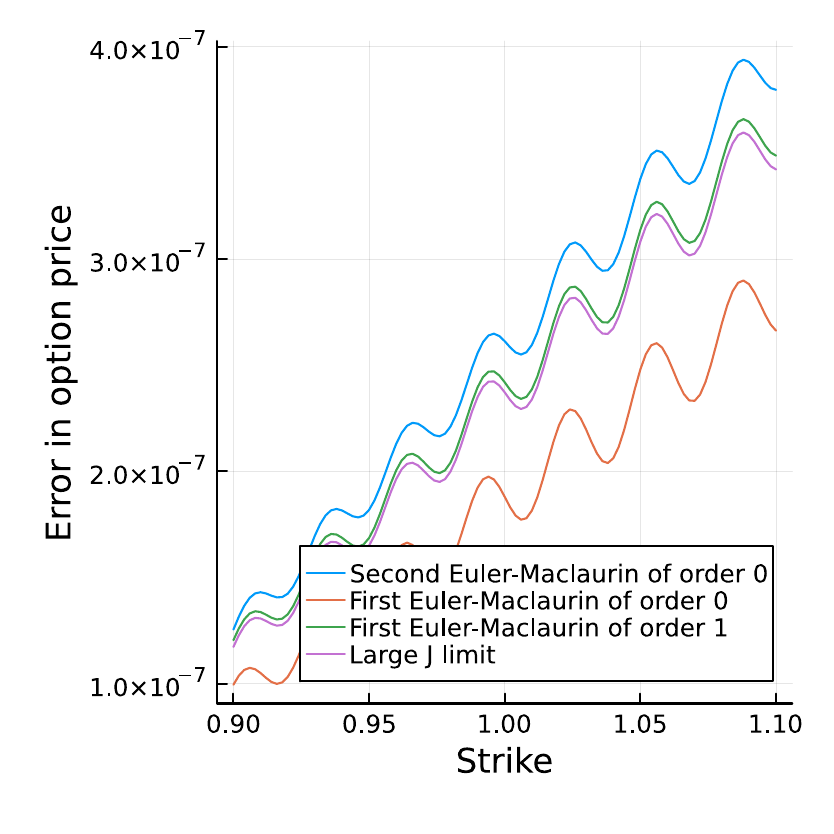}}
		}
		\caption{\label{fig:price_error_by_density_set2}Error in density coefficients and vanilla option prices for a range of strikes, using $m=6, L=8, J=2^4$ for the Heston parameters of Set 2.}
	\end{figure}
	
We may deduce that the mid-point rule is slightly better since it 	involves $J$ instead of $J-1$ points. In reality, the first point of the trapezoidal rule is $\phi(0) = 1$, and is thus not
more costly. Furthermore, the FFT is simplified and does not involve $J$ evaluations of the complex exponential function with modulus one .  
More importantly, the trapezoidal rule allows to reuse the previous characteristic function evaluations when  $J$ is doubled, which may be particularly important in the procedure used to choose the SWIFT parameters. 

In addition we may use more accurate quadrature rules. For example, the first Euler-Maclaurin formula with first derivative correction reads
\begin{align}
	c_{m,k} &=  2^{\frac{m}{2}+1} \Re\left\{\int_0^{\frac{1}{2}} \hat{f}\left(2^{m+1} \pi t\right) e^{2i \pi k t} \diff t\right\}\label{eqn:density_parseval}
		\\&=\frac{2^{\frac{m}{2}}}{J} \sum_{j=0}^J w_j \Re\left\{ \hat{f}\left(\frac{\pi j}{J}2^m\right) e^{i \frac{\pi j}{J} k}\right\} - \frac{2^{\frac{m}{2}}}{12J^2 }\Re\left\{
		i\pi k \left(\hat{f}(2^{m}\pi)e^{i \pi k }-\hat{f}(0)\right)+ 2^{m}\pi\left(\hat{f}'(2^m\pi)e^{i \pi k }-\hat{f}'(0)\right)   \right\}   
	\end{align}
with weights $w_0=w_j=\frac{1}{2}$ and $w_j=1$ for $j=1,...,J-1$.
From Figure \ref{fig:price_error_by_density_set2}, the accuracy of the calculation of the density coefficients $c_{m,k}$ is improved, but the error in the price of European options not reduced.  

The Simpson quadrature requires double the number of evaluations of the characteristic function (if we halve $J$, the error may explode) and again does not translate into a more precise price.

\section{Corner cases}
Many of the corner cases below come from the exhaustive parameter set of \citet{andersen2019heston}, used to benchmark their pricing method.

In spite of the more refined initialization, the SWIFT method has issues on many of the same examples as the COS method.
With the Heston parameters $v_0=10^{-4}, \kappa=0.1, \theta= 0.25, \rho=0.95, \sigma=3.0, T=10$,  a $\epsilon_m = 10^{-5}$ leads to $m=7$ while $\epsilon_m=10^{-6}$ leads to $m=8$.
Then the procedure of \citet{leitao2018swift} leads respectively to $J=2^{18}$ and $J=2^{19}$ for $\epsilon_f=10^{-8}$. The overall accuracy is indeed good (while COS with $M=200$ and $L=8$ leads to bad prices (it requires several thousand points and a larger $L$) but the performance is inadequate: ignoring the initialization procedure, $J$ corresponds to the number of distinct calls of the characteristic function.

With $v_0=10^{-4}, \kappa=0.01, \theta= 1.0, \rho=-0.95, \sigma=3.0, T=10$, and the same accuracy levels, we obtain $m=9$ and $J >2^{20}$. The choice $m=7$ would be sufficient (Table \ref{tbl:swift_corner2}), but it still implies $J=2^{18}$, which is huge. The guess for the interval $[-c,c]$ is good using $L=8$ and four cumulants. It leads to a very large but adequate $c=1094.6$. With $L=1$, the initialization procedure leads to $c=1101.4$, very close. The COS method attain a good accuracy with those settings only for $M \geq 2^{16}$ points, which is also huge and its accuracy decreases for smaller values of $L$, thus confirming the validity of the SWIFT initialization procedure. 

\begin{table}[h]
	\caption{Error in the price of a vanilla call option for different settings of the SWIFT method and a range of option strikes. The reference prices, obtained by the technique of \citet{andersen2019heston} are respectively 3.032277336306425,		3.2085075362598046,
		10.087170493728104,		100.00002701432814,		900.0000000000015,		9900.0 for an asset with forward price $F=100$. \label{tbl:swift_corner2}}
	\centering{
\begin{tabular}{llllrrrrrr} \toprule
	$m$ & $L$ & $\epsilon_f$ &  $J$ & \multicolumn{6}{c}{Strike} \\ \cmidrule(lr){5-10}
	& & & & 100.0001 & 101 & 110 & 200 & 1000 & 10000 \\\midrule
	9 & 8 & $10^{-8}$ & $2^{21}$ & -3.17e-07& -3.20e-07& -3.48e-07& -6.34e-07& -3.17e-06& -3.17e-05\\
	9 & 8 & $10^{-4}$ &  $2^{19}$ & -4.03e-03 & -4.07e-03& -4.43e-03& -8.05e-03& -4.03e-02& -4.03e-01 \\
	8 & 8 & $10^{-4}$ & $2^{18}$ & -4.03e-03& -4.07e-03& -4.43e-03& -8.05e-03& -4.03e-02 & -4.03e-01\\
	8 & 8 & $10^{-6}$ & $2^{19}$ & -4.37e-05& -4.38e-05& -4.82e-05& -8.74e-05& -4.37e-04& -4.37e-03 \\
8 & 8 & $10^{-8}$ &$2^{20}$ &	-1.08e-05& -1.06e-05& -1.21e-05& -2.17e-05& -1.09e-04& -1.09e-03\\
7 & 8 &  $10^{-8}$ & $2^{19}$& 2.49e-05& -1.11e-04& -1.53e-05& -1.12e-05& -2.88e-05& -1.16e-04\\
	7 & 8 & $10^{-6}$ & $2^{18}$ & -1.48e-05& -1.51e-04& -5.90e-05& -9.06e-05& -4.26e-04& -4.09e-03\\
	6 & 8 & $10^{-6}$ &$2^{17}$ &  1.20e-03& 1.41e-03& -3.01e-05& -3.64e-04& -1.52e-04 & -6.96e-03\\
	5 & 8 & $10^{-6}$ & $2^{16}$ & -1.62e-02& -2.95e-02& 4.09e-03& 1.89e-03& -3.22e-03 & -1.38e-02\\ \bottomrule
\end{tabular}}
\end{table}

The cause of the inefficiency is directly related to the very large interval $[-c,c]$ required on those examples. A smaller $c$ would decrease accuracy, and such a large $c$ requires a huge number of points.

Furthermore, in the end, the method becomes much slower than COS.

\section{Conclusion}
There are not many improvements to propose on the SWIFT method as described in  \citep{maree2017pricing, leitao2018swift}: more precise approximations of sinc do not translate into a more precise option price. The initialization procedure of \citet{leitao2018swift} is more agnostic to the choice of initial range.  And in this regard, the SWIFT method is more robust than the COS method. We also proved that the expansions were converging with a slightly smaller $J$ than the choice presented in \citep{ortiz2016highly} and subsequent papers. 

The use of the trapezoidal rule is preferable to the use of the Vieta based expansion, in order to reuse points when the number of points is doubled.

Similarly to the COS method, the SWIFT method requires too many points to be practical on many corner cases.  This may however not be a problem for the calibration of stochastic volatility models or Levy models, if a good initial guess and a proper error measure are used. 


\externalbibliography{yes}
\bibliography{swift_notes_revisited.bib}
\appendixtitles{no}

\end{document}